\documentclass{IOS-Book-Article}

\usepackage{mathptmx}
\usepackage{soul}\setuldepth{article}
\usepackage{graphicx}
\usepackage[T1]{fontenc}
\def\hb{\hbox to 11.5 cm{}}

\usepackage{hyperref}
\hypersetup{colorlinks,allcolors=black}
\usepackage[capitalise]{cleveref}
\usepackage{tabularx}
\usepackage{listings}
\usepackage{xcolor}

\newcommand{\mt}[1]{\textsf{#1}}
\newcommand{\str}[1]{\textbf{\mt{\guillemotleft {#1}\guillemotright}}}

\input{resources/sparql.sty} % SPARQL syntax highlighting
\input{resources/owl.sty} % OWL syntax highlighting
\input{resources/ocl.sty} % OCL syntax highlighting

\crefname{lstlisting}{Listing}{Listings}
\Crefname{lstlisting}{Listing}{Listings}

\begin{document}

\pagestyle{headings}
\def\thepage{}
\begin{frontmatter}              % The preamble begins here.

%\pretitle{Pretitle}
\title{Towards a conceptual model for the FAIR Digital Object Framework}

\markboth{}{FOIS 2023\hb}
%\subtitle{Subtitle}

\author[A, B]{\fnms{Luiz Olavo} \snm{Bonino da Silva Santos}%\orcid{0000-0002-1164-1351}%
\thanks{Corresponding Author: Luiz Olavo Bonino da Silva Santos, l.o.boninodasilvasantos@utwente.nl.}},
\author[A]{\fnms{Tiago Prince} \snm{Sales}%\orcid{0000-0002-5385-5761}
}, \author[A]{\fnms{Claudenir M.} \snm{Fonseca}%\orcid{0000-0003-2528-3118}
}
and
\author[A]{\fnms{Giancarlo} \snm{Guizzardi}%\orcid{0000-0002-3452-553X}
}

\runningauthor{Bonino da Silva Santos et al.}
\address[A]{Semantics, Cybersecurity {\usefont{OT1}{cmtt}{m}{it} \&} Services (SCS), University of Twente, The Netherlands}
\address[B]{Biosemantics group, Leiden University Medical Center, The Netherlands}

\begin{abstract}
The FAIR principles define a number of expected behaviours for the data and services ecosystem with the goal of improving the findability, accessibility, interoperability, and reusability of digital objects. A key aspiration of the principles is that they would lead to a scenario where autonomous computational agents are capable of performing a ``self-guided exploration of the global data ecosystem,'' and act properly with the encountered variety of types, formats, access mechanisms and protocols. The lack of support for some of these expected behaviours by current information infrastructures such as the internet and the World Wide Web motivated the emergence, in the last years, of initiatives such as the FAIR Digital Objects (FDOs) movement.  This movement aims at an infrastructure where digital objects can be exposed and explored according to the FAIR principles. In this paper, we report the current status of the work towards an ontology-driven conceptual model for FAIR Digital Objects. The conceptual model covers aspects of digital objects that are relevant to the FAIR principles such as the distinction between metadata and the digital object it describes, the classification of digital objects in terms of both their informational value and their computational representation format, and the relation between different types of FAIR Digital Objects. %To do that, we propose a model that puts together theoretical notions related to a Four-Category Ontology, a minimalistic Multi-Level Modelling mechanism, and some basic aspects of an Ontology of Information Artefacts. 
\end{abstract}

\end{frontmatter}
\markboth{FOIS 2023\hb}{FOIS 2023\hb}

\section{Introduction}

The evolution of the informatics infrastructure has happened so far in incremental and complementary steps. Whenever the challenges presented in one step are addressed, a new set of possibilities emerge and, with them, new challenges. From the challenges of interconnecting computers in a network to interconnecting different networks (the internet) and interlinking networked resources (the World Wide Web), one new layer of interoperability facilitation leverages from the previous one. 

%In the digital realm, we constantly interact with different types of entities, or objects. Ultimately, these objects are represented by sets of bits stored in a digital medium. For instance, all documents present in a computer such as text files, images, video, audio, etc., as well as software programs are represented as bit sequences and stored in a computer's hard drive, solid-state drive or other types of storage medium. The totality of bits stored in the medium is divided into different units, representing different types of objects and distinguished based on their different informational value. For instance, the bits sequence representing a text document, which contains an essay constitutes a different informational unit than another bits sequence representing the text editor software.
In the digital realm, we constantly interact with different types of entities, or objects. The totality of bits in a computer's storage medium is divided into different units, representing different types of objects and distinguished based on their different informational value. For instance, the bits sequence representing a text document, which contains an essay constitutes a different informational unit than another bits sequence representing the text editor software.

The FAIR principles \cite{Wilkinson2016} brought a vision of a global ecosystem of dynamically interoperable objects, such as data, services and computing capacity. This vision includes computational agents that are ``\emph{capable of autonomously and appropriately acting when faced with the wide range of types, format, and access mechanisms/protocols that will be encountered in their self-guided exploration of the global ecosystem}''.

This envisioned scenario requires that the exploring computational agents are provided with enough actionable information about the encountered objects so that they can automate as much as possible the object's discovery, access, interoperation and reuse. The provision of information about digital objects entails not only that the information is made available to the agents, but it is also required that the infrastructure, participating systems and services expose a number of behaviours for the interactions involved in this provision to be supported. Therefore, to fully realise that vision, we need an infrastructure capable of supporting the manipulation of digital objects according to the requirements defined by the FAIR principles as well as definitions of the expected behaviours of the participating systems, applications and services.

These challenges to apply the FAIR principles in the current digital communication infrastructure are the main motivators of the work on the FAIR Digital Object Framework (FDOF). The FDOF is a framework designed to support the presentation of informational objects in a digital environment. As the name suggests, the FDOF is inspired by the notion of Digital Objects introduced in \cite{kahn_framework_2006} and extended it to comply with the additional requirements derived from the FAIR principles. 

The FDOF aims at defining features to provide infrastructural support for the FAIR principles. This set of features added to the underlying communication infrastructure is concentrated in requirements for a predictable identifier resolution behaviour, a retrieval mechanism for object metadata from the object's identifier and an object typing system.

To define the entities and relations included in the framework we have defined an ontology-based conceptual model named FAIR Digital Object Framework Conceptual Model (FDOF-CM), which is the focus of this paper. The remainder of the paper is organised as follows: \cref{sec:research-baseline} introduces the FAIR Digital Object Framework and discusses the ontological foundations used in our proposed conceptual model; \cref{sec:fdof-ontology} presents the FDOF Conceptual Model detailing its main entities and relations; \cref{sec:fdof-evaluation} evaluates the FDOF-CM by deriving an OWL ontology which is validated in a set of relevant use cases; \cref{sec:related-work} discusses our related work; and \cref{sec:conclusions} presents final considerations and future work.

\section{Research Baseline}
\label{sec:research-baseline}

The focus of this paper is the conceptual model defining and describing the entities and relations present in the FAIR Digital Object Framework. Below we provide a research baseline of the work on the FDOF and the ontological notions used in the FDOF-CM.

\subsection{FAIR Digital Object Framework}
\label{sec:fdof-background}

In an increasingly complex digital environment automation becomes a necessity and to support increasing levels of automation, proper identification, and classification of different digital objects become relevant as well as the qualification of the relations between them. This brings additional requirements to the digital ecosystem, which are not properly covered by the internet or the World Wide Web. The latter, for instance, is based on a network of digital objects linked through unqualified links, the web links, i.e., the links between web resources are not semantically defined.

In the seminal paper where the FAIR principles have been introduced \cite{Wilkinson2016}, the authors list enabling capabilities that autonomously-acting computational data exploring agents acquire from information provided by the underlying infrastructure when facing newly discovered digital objects. These capabilities are: \emph{``(1) identify the type of object (with respect to both structure and intent), (2) determine if it is useful within the context of the agent's current task by interrogating metadata and/or data elements, (3) determine if it is usable, with respect to license, consent, or other accessibility or use constraints, and (4) take appropriate action, in much the same manner that a human would.''}

These required capabilities can be converted into the following questions that the framework should support and enable both computational agents and humans to answer: (Q1) What type of object does this encountered identifier identifies? (Q2) How can I get more information (e.g., metadata) about this object? (Q3) How can this object be manipulated, by whom, and under which conditions?  

%\begin{itemize}
 %   \item What type of object does this encountered identifier identifies?
  %  \item How can I get more information (e.g., metadata) about this object?
   % \item How can this object be manipulated, by whom, and under which conditions?
%\end{itemize}

Given their different nature, digital objects can be manipulated in different ways. The manipulation of digital objects occurs in a combination of computational agents and the underlying technological infrastructure. The interactions happen considering at least two aspects of the digital objects, their \textit{representation or materialisation format} and their \textit{informational content}. The former distinguishes ways of manipulating different materialisation formats such as an executable program, a PDF file, or a video file. While the executable program can be run on a given operating system, we can ``open'' the PDF file for reading or editing using specific supporting software and we can play the video file to watch the animated sequence also using a video player software. In other words, the possibility of interacting with a given digital object depends on the ability of the computational agent or technological infrastructure to handle the object's encoding format.

The informational aspect of the digital objects is dealt with by inspecting the content of the bit sequences considering their different informational roles, e.g., does a given bit sequence represent a structured dataset, a software code, a natural language text document, or images? Many materialisation formats are directly related to specific informational roles such as the MP4 format associated with video sequences and the XML format related to structured data. But in other cases, this relation is not so clear. For instance, we can materialise digital images, text, and structured data using PDF files. In these cases, knowing the materialisation format is not enough to clarify the informational nature of the digital object. If we want to increase the autonomy of computational agents in their handling of digital objects, we should also provide information about each object's informational value besides its materialisation format.

In order to support these questions and challenges the FDOF should rely on a set of clear and unambiguous definitions of the involved concepts and their relations as well as determine expected behaviours from the underlying infrastructure and participating computational agents. The proposed conceptual model discussed in section \ref{sec:fdof-ontology} aims at providing these clear and unambiguous conceptualizations related to the FDOF. Our focus here is, therefore, the presentation, description, and discussion of the FDOF conceptual model. For a more in-depth explanation of the framework itself, its motivation, and historical evolution, the FDOF documentation \cite{FDOFDoc:2021} should be consulted.

\subsection{Ontological Foundations}
\label{sec:ontouml}

Conceptual models are artefacts developed to provide semantic clarification of a subject domain. One approach to building conceptual models is to employ foundational ontologies that assist the modeller in consistently representing fundamental aspects of the domain. We refer to this approach as ontology-driven conceptual modelling (ODCM) \cite{verdonck2015ontology}. We use in this paper an ODCM extension of UML called OntoUML. OntoUML enriches UML's class diagram language with constructs (e.g., classes and relations) that reflect the ontological categories of the Unified Foundational Ontology (UFO) \cite{guizzardi2005ontological}. Additionally, OntoUML defines \textit{semantically-motivated syntactical constraints} designed to ensure that valid models represent sound UFO ontologies.

OntoUML uses special labels called stereotypes to decorate classes and relations that express certain ontological categories from UFO. The stereotyped classes, hereby referred to as types, reflect two orthogonal classifications criteria from UFO: (i) classifications based on the ontological nature of the type's instances (e.g., objects, individualized properties, abstract values); and (ii) classifications based on modal characteristics of how the type applies to its instances (e.g., necessarily or accidentally). We now summarize the stereotypes later employed in \cref{sec:fdof-ontology}.

The UFO ontology employs the notion of identity principle to refer to the best-suited criteria to necessarily classify the different sorts of concrete individuals in a domain. For example, a modeller could select the type \mt{Organization} as the type providing the identity principle of its instances within a taxonomy ranging from \mt{Entity} to \mt{Professional Limited Liability Organization}. In OntoUML, these \textit{ultimate sortals} (also known as \textit{kinds}) are represented by a set of stereotypes that identify both identity principle and the ontological nature of their instances. The stereotype \str{kind} identifies ultimate sortals whose instances are objects in the domain (e.g., \mt{Organization}, \mt{Person}, \mt{Car}), while \str{relator} identifies ultimate sortals whose instances are individualized relational properties (e.g., \mt{Marriage}, \mt{Employment}, \mt{Enrollment}).
The stereotype \str{mediation} decorates the associations between these types of relational properties and the types of objects on which their instances are existentially dependent (i.e., their \emph{bearers}).

While ultimate sortals uniquely and necessarily classify the entities following the identity principle it provides, other types of concrete individuals either classify individuals following a common principle or classify individuals of a variety of principles. Classes decorated by \str{subkind} or \str{role} inherit a principle provided by an ultimate sortal, where the former necessarily classifies its instances (e.g., \mt{Limited Liability Organization}, \mt{Sports Car}), and the latter does so accidentally based on some relational property (e.g., \mt{Hired Organization}, \mt{Rented Car}). Classes decorated with \str{category} or \str{roleMixin} are similar to subkinds and roles in classifying instances either necessarily (e.g., \mt{Agent}, \mt{Vehicle}) or accidentally (e.g., \mt{Acting Agent}, \mt{Damaged Vehicle}), but these classify concrete individuals following different identity principles, i.e., instances of multiple kinds.

% We also employ two additional stereotypes from OntoUML, \str{abstract} and \str{type}.

Types decorated by \str{referenceStrucutre} and \str{referenceRegion} \cite{albuquerque2013ontological} classify abstract individuals representing structured conceptual spaces (e.g., the three-dimensional \mt{Colour Space}) and individualized regions within these spaces (e.g., the \mt{Colour Value} \textit{Crimson-H348°-S91\%-B86\%}) respectively. Types decorated by \str{type} are higher-order types whose instances are themselves other types that are part of the subject domain \cite{fonseca2022types}, thus, allowing modellers to describe them and their (higher-order) properties much in the same way one describes regular individuals. Associations decorated with \str{instantiation} reify the notion of instantiation in the model connecting these higher-order types to their so-called \textit{base types}, i.e., the super types of their instances, providing information about the characteristics their instances must inherit and forming the powertype pattern. Furthermore, associations decorated with \str{historicalDependence} indicate historical dependencies of instances of the association's target on instances of the source (e.g., in the case of creation relations). Finally, the relation of \textit{redefinition} of association ends from UML has also been employed in \cref{sec:fdof-ontology}. This allows the modeller to bind association ends connected to specialised types, to the association ends connected to their parents, expressing a refinement of their parents' association with additional constraints. %The association end subsetting represents an association end that is included in the set of relata of the subsetted one. For example, we can express that a faculty department that \textit{has members} that are senior researchers subsets a faculty department that \textit{has members} that are researchers in general, in both cases the \textit{has members} association being the same. The redefinition, on the other hand, has more strict semantics where instances of the type connected through the redefining end can only be involved through that association and respecting its constraints. We could represent, for instance, that a faculty department with a \textit{has lead researcher member}, with this association end redefining entirely that association that could relate to a lead researcher.

To facilitate the readability of the ontological natures captured in OntoUML classes' tagged values, we use a colour coding to denote the ontological nature of their instances, with red denoting objects, green denoting relational properties, purple denoting types, and white denoting abstracts.

% TODO: add \str{mediation}, \str{historicalDependence}, and \str{instantiation}.

\section{The FAIR Digital Object Framework Conceptual Model}
\label{sec:fdof-ontology}

The goals of the FAIR Digital Object Framework Conceptual Model are to define the main concepts of the framework and how they relate, and to provide the basis for an ontology of these concepts and relations. This ontology should be used to semantically annotate the elements of the framework in metadata records, semantic data and metadata models, or whenever the representation language used in the digital object allows the inclusion of semantic annotations, e.g., RDF, OWL.

To facilitate understanding, we have divided the FDOF-CM into three parts: object identification, basic typing system, and the distinction between metadata and the object it describes.
\subsection{FDO Identification}
\label{subsec:FDO-ID}

Proper identification of different digital objects is a central point in the FAIR principles. Three principles explicitly mention identifier, namely: \emph{``F1. (meta)data are assigned a globally unique and persistent identifier''}, F3 (\emph{``metadata clearly include the identifier of the data it describes''}) and A1 (\emph{``(meta)data are retrievable by their identifier using a standardized communication protocol''}).

\begin{figure}[ht]
\centering
\includegraphics[width=\linewidth]{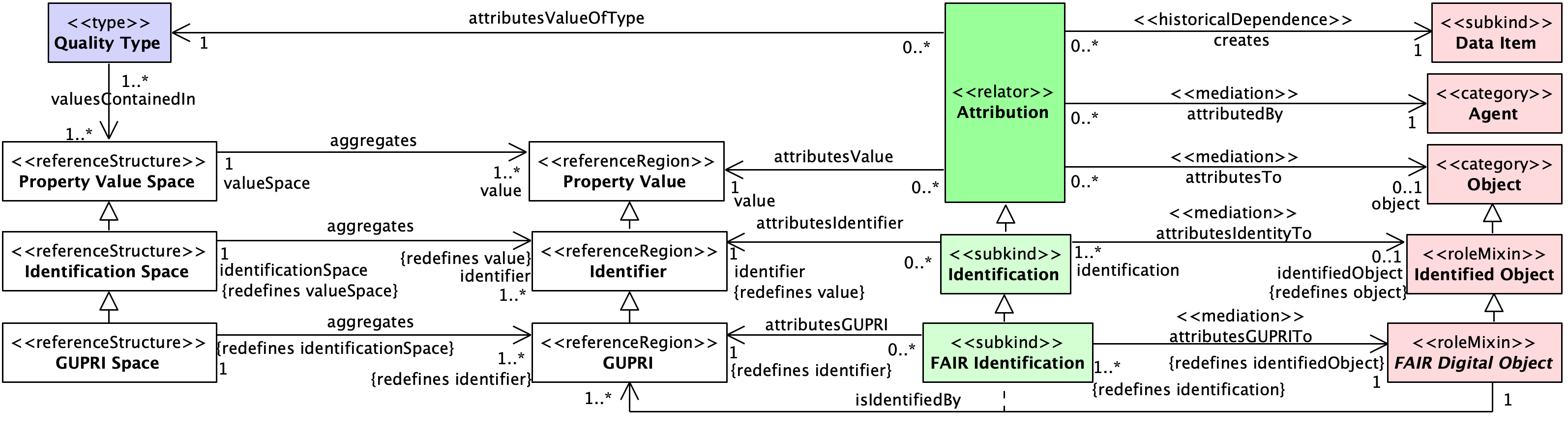}
\caption{FDOF - Identification and attribution}
\label{fig:fdof-attribution-and-identification}
\end{figure}

Figure \ref{fig:fdof-attribution-and-identification} presents a fragment of the FDOF Conceptual Model where a \mt{FAIR Digital Object} is defined as a specialisation of \mt{Identified Object}. However, differently from an \mt{Identified Object} that is generically identified by an \mt{Identifier}, \mt{FDOs} are identified by an specialisation of \mt{Identifier} that is globally unique, persistent and resolvable (\mt{GUPRI}). The refinement from \mt{Identified Object} to \mt{FAIR Digital Object} and from \mt{Identifier} to \mt{GUPRI} is a consequence of the requirements of the FAIR principles F1 (for the globally uniqueness and persistence) and A1 (for being resolvable) that a FAIR Digital Object not only needs to be identified by an identifier but this identifier must guarantee global uniqueness (single reference), persistence (rigid designation) and be resolvable (retrievable from its identifier).

The model in Figure \ref{fig:fdof-attribution-and-identification} defines that \mt{Identifiers} are \mt{Property Values} within a specialisation of \mt{Property Value Space} named \mt{Identification Space}. The \mt{Identification Space} represents all possible values (identifiers) of that identification system. One example is that the identification space for Unified Resource Identifiers (URIs) is the set of possible values for URIs as defined in \cite{URI} following the pattern of a scheme, a colon (:), a single or double forward slash (/ or //), an authority, a path, a query and a fragment.

Another aspect covered in this FDOF-CM identification fragment is the reification of the \mt{Identification} relation between \mt{Identifiers} and \mt{Identified Objects}, and more specifically the \mt{FAIR Identification} relation between \mt{GUPRIs} and \mt{FDOs}, that are made by \mt{Agents}. In this way, we can capture the information of who assigned an identifier to a given FDO, adding to the object's provenance.

\subsection{FDOF Basic Typing System}
\label{subsec:FDOF-types}

%Classification is an essential mechanism for communication. Natural languages make extensive use of it by means of nouns. Common nouns such as chair, dog or mountain do not refer to specific individual entities but denote a class of objects that share a set of characteristics. These common nouns allow us to classify a piece of furniture as a chair and not as a bed, an individual canine animal as a dog or a given geographic feature as a mountain. Similarly, types are used in conceptual modelling as entities that define a set of generic characteristics shared by their instances. The definition of this set of shared characteristics allows us to simplify the universe of discourse by aggregating a number of individuals under one grouping entity. %As stated in \cite{Porello_Guizzardi_2017}, \emph{``if each individual entity needed a distinct name, our language would be staggeringly complex and communication virtually impossible''}.

Types are taken here to be aggregations of properties that are used to characterise their members. Commonly, the choice is made to fulfil a particular goal or use. In the realm of digital objects, different typing systems have been used for different purposes, mainly related to creating abstractions for the sets of bits stored in computer memory. For instance, most operating systems use the abstraction of files to organise groups of bits whereas data sets, potentially spanning over several files or as the content of database management systems, are commonly used in data science activities.

On the internet, types of digital objects are often associated with Media Types (formerly known as MIME types) \cite{IANA_MediaTypes}. By looking at the hundreds of types defined by the Internet Assigned Numbers Authority (IANA) as Media Types, one cannot avoid noticing some inconsistency in the classification criteria when taking into consideration the nature of the defined entities. In some cases, the media type represents a given file encoding format, e.g., \emph{application/dicom}, \emph{audio/mpeg}, \emph{image/png} or \emph{video/h264}. In other cases, the media type represents not only a generic encoding format but how the content is organised in a particular structure, e.g., the IANA media type \emph{application/alto-directory+json} defines both the encoding format (\emph{JSON}) and a specific structure of key-value pairs following the Application-Layer Traffic Optimization (ALTO) protocol.

We consider that this typing system based only on the encoding format is insufficient to support the levels of automation envisioned by the FAIR principles. For instance, when trying to classify a given photo as an FDO, should the type of the FDO represent the image captured in the photo, the encoding format of the photo file or what is depicted in the photo? We argue that we should consider these three aspects as part of the characterisation of a given FDO, namely, \emph{(i)} the encoding format, \emph{(ii)} the type of the object regarding its informational content and \emph{(iii)} the entity(ies) represented by the object. The motivation for the argument is that a computational agent benefits from information in these three levels of abstraction to improve its machine-actionability as required by the FAIR principles. The encoding format helps determine how the digital object can be processed by the computing infrastructure (e.g., which parser or reader should be used). The type of the object considering its informational content helps differentiate, for instance, images from text, video, applications, etc. Finally, the reference to the entities represented by the object provides more information about what the object represents.

\begin{figure}[ht]
\centering
\includegraphics[width=.65\linewidth]{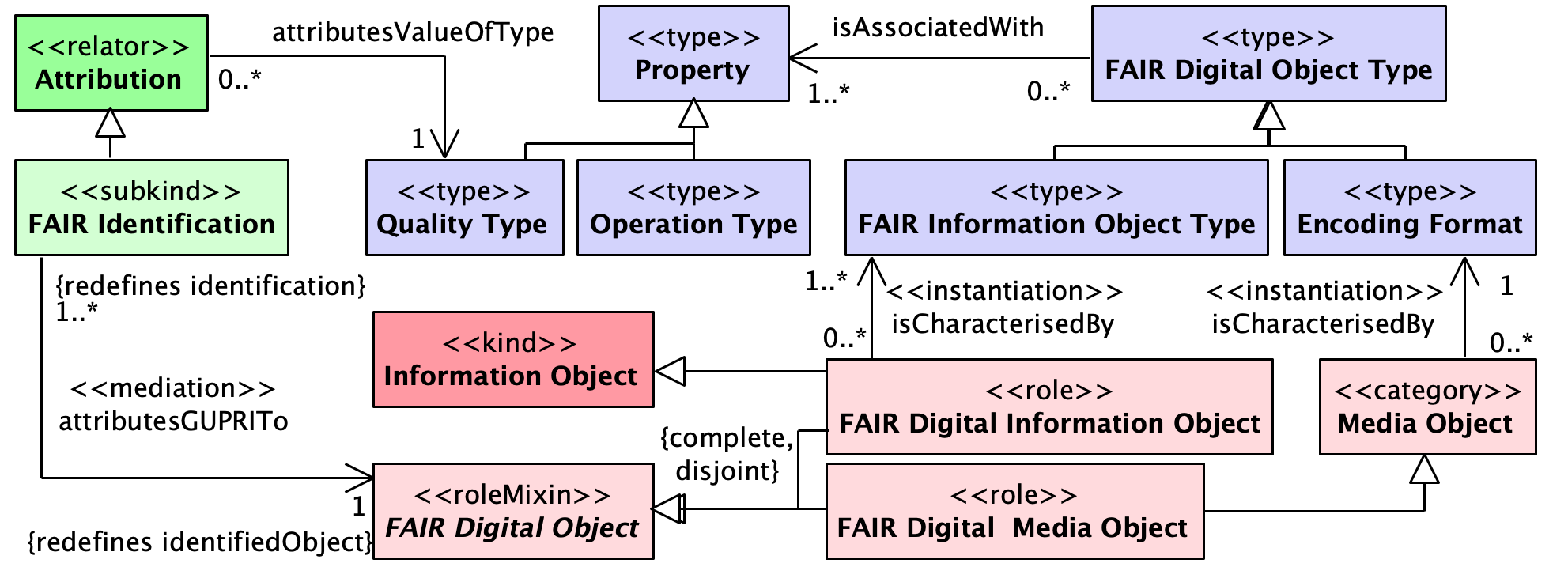}
\caption{FDOF - FAIR Characterisation}
\label{fig:fdof-characterisation}
\end{figure}

\cref{fig:fdof-characterisation} depicts a fragment of the FDOF-CM containing the main classes of the FDOF basic typing system. In this model, a \mt{FAIR Digital Object} can be either a \mt{FAIR Digital Information Object} (FDIO) or a \mt{FAIR Digital Media Object} (FDMO). The \mt{FDIO} class represents FDOs as units of information. As shown in this model, \mt{Information Object} defines a kind, which principle of identity is defined by the identity of information content. Every \mt{FDIO} is an \mt{Information Object} materialized by at least one \mt{FDMO}. The FDMO is bit sequence\footnote{The kind \mt{Bit Sequence} in Fig. \ref{fig:fdof-metadata} defines a self-evident principle of identity inherited by FDMOs.}, which (as all \mt{Media Objects}) encodes characteristics that are proper to the media type (representing here by \mt{Encoding Format}). Every \mt{FDMO} encodes the information content of a \mt{FDIO}.  In the photo example, John's birthday party photo (an instance of \mt{FDIO}) is materialised by an object named \emph{JohnBDayPhoto1.jpg}, which is an instance of \mt{FDMO} and has the encoding format \emph{image/jpg} (instance of \mt{Encoding Format}). We could also have the same \mt{FDIO} materialised by another \mt{DFMO} instance named \emph{JohnBDayPhoto2.heic} with the encoding format \emph{image/heic}. In this example, we have three FDOs involved, one \mt{FAIR Digital Information Object}, John's birthday party photo, and two \mt{FAIR Digital Media Objects}, \textit{JohnBDayPhoto1.jpg} and \textit{JohnBDayPhoto2.heic}.

In general, a given FDO is characterised by one sub-type of \mt{FAIR Digital Object Type} (FIOT). FIOTs aggregate the set of properties that characterise information objects of that type (see \textit{isAssociatedWith} relation in Fig. \ref{fig:fdof-characterisation}) . For instance, if we have the photo of John's birthday party, we can classify this object as a \mt{FDIO} having the information object type \emph{Photo}. This \emph{Photo} \mt{FIOT} aggregates a number of properties that characterise photos, e.g., the event depicts, the participants, location, date and time, etc. However, as any \mt{Object} here, this photo can have a number of \mt{attributions} referring to other \mt{Identified Objects} (e.g., an attribution \textit{participant} can point to an \mt{Identified Object} of the object type Person, an attribution \textit{location} can point to a \mt{property value} in a geo-coordinate space).  

%In general, a given FDO is characterised by one sub-type of \mt{FAIR Digital Object Type} (FIOT). More specifically, an FDO of type \mt{FAIR Digital Information Object} is characterised by one \mt{FAIR Digital Information Object Type} (FDIOT). FDIOTs aggregate the set of properties that characterise information objects of that type (see \textit{isAssociatedWith} relation in Fig. \ref{fig:fdof-characterisation}) . For instance, if we have the photo of John's birthday party, we can classify this object as a \mt{FDIO} having the information object type of \emph{Digital Photo}, which is an instance of \mt{FDIOT}. This \emph{Digital Photo} \mt{FDIOT} aggregates a number of properties that characterise photos, in general, e.g., who took the photo, location that has been taken, date and time, aperture setting, shutter setting, entities depicted, etc., as well as properties that are reserved to digital photos.

\subsection{Metadata}
\label{sec:FDO-metadata}

\begin{figure}[ht]
\centering
\includegraphics[width=\linewidth]{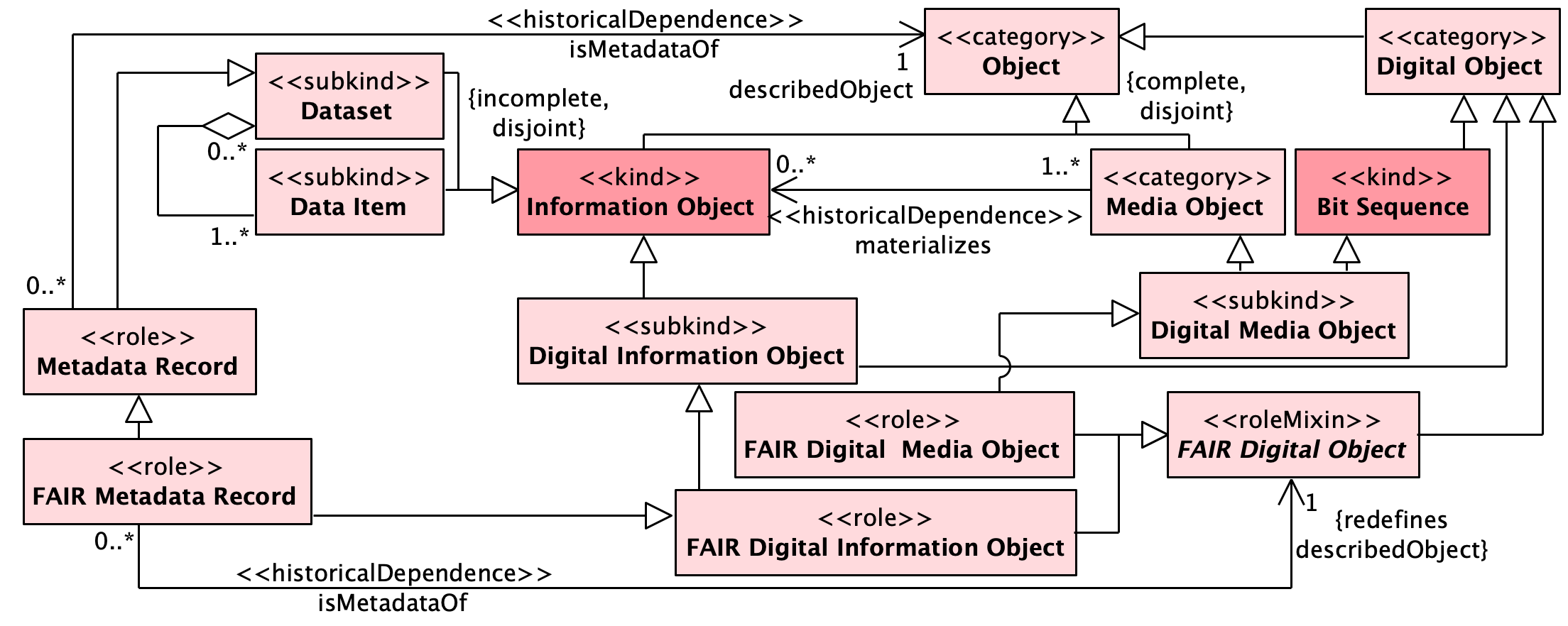}
\caption{FDOF - Metadata Core}
\label{fig:fdof-metadata}
\end{figure}

One important element of the FAIR principles is the role attributed to metadata as responsible for describing a given object (principle F2). Many of the FAIR principles  indicate that they apply to both metadata and other types of objects (principles F1, F4, A1, I1, I2, I3, R1.1, R1.2 and R1.3). Because of this, the FDOF-CM defines the \mt{FAIR Metadata Record} (FMR) as a specialisation of \mt{FAIR Digital Information Object} with the role of describing \mt{FAIR Digital Objects}, as depicted in \cref{fig:fdof-metadata}. 

According to our model, the \mt{FMR} also specialises \mt{Dataset}, which aggregates \mt{Data Items}. In this way, a \mt{FMR} is a sort of dataset information object which contains a number of data items representing the attributions of values to properties that are used to describe a particular object. As an specialisation of \mt{FDIO}, the \mt{FMR} is characterised by a \mt{FAIR Information Object Type}. The FDOF-CM supports the possibility of defining specialisations of \mt{FAIR Information Object Type} that associate the sets of properties that can (optional) or should (mandatory) be used to characterise the metadata record. In an implementation, these definitions can be realised in terms of metadata schemas associated with given types of FDOs.

The class \mt{FDO} is \textit{abstract} (in the UML sense), i.e., cannot be directly instantiated. Instances of \mt{FDO} must be either instantiate \mt{FAIR Digital Information Object} or \mt{FAIR Digital Media Object}. Consequently, both \mt{FDIOs} and \mt{FDMOs} should be described by metadata records.

However, the requirement that an \mt{FDO} must be described by a metadata record and the fact that a \mt{FMR} is an specialisation of \mt{FDO} creates the possibility of an infinite loop of metadata records requiring being described by meta-metadata records. To avoid this loop and to better reflect the reality, the FDOF-CM includes the following constraint rule: $\mt{FAIRDigitalObject}(x) \land \neg\mt{FAIRMetadataRecord}(x) \rightarrow \exists y \, (\mt{FAIRMetadataRecord}(y) \land \mt{isMetadataOf}(y,x))$. This rule redefines the zero-to-many cardinality of the \mt{isMetadataOf} relation between \mt{FMR} and \mt{FDO} to require at least one metadata record to describe the FDO unless the FDO is an instance of \mt{FMR}. With this constraint rule, the model still allows metadata records to have their own metadata records but do not mandate it.

%\begin{itemize}
 %   \item[(\textbf{a1})] $\mt{FAIRDigitalObject}(x) \land \neg\mt{FAIRMetadataRecord}(x) \rightarrow \\ \hspace*{1cm} \exists y \, (\mt{FAIRMetadataRecord}(y) \land \mt{isMetadataOf}(y,x))$
%\end{itemize}

% \begin{lstlisting}[language=SPARQL,caption={FAIR Digital Object metadata constraint.},label={owl:ocl-metadata}]
% context FAIRDigitalObject inv requiredMetadata: 
% not self.oclIsTypeOf(FAIRMetadataRecord) implies self.hasMedatada->size() >= 1
% \end{lstlisting}
 
As the model fragment depicted in \cref{fig:fdof-metadata} indicates, we make a distinction between digital and non-digital objects. This supports the possibility of having, for instance, FAIR Metadata Records describing physical objects. However, in this paper, we focus only on digital objects. In fact, in the FDOF-CM we have generic elements such as \mt{Identified Object}, \mt{Information Object}, \mt{Media Object} and \mt{Metadata Record}. But what makes them FAIR? According to the FDOF, an \mt{Identified Object} becomes a \mt{FDO}, regarding identification, when it is identified by a \mt{GUPRI} and not any other type of identifier, as depicted in \cref{fig:fdof-attribution-and-identification}. Similarly, a generic \mt{Metadata Record} becomes a \mt{FMR} when is assigned with a \mt{GUPRI} and includes the identifier of the object it describes (per FAIR principle F3) using the relation \mt{isMetadataOf} (\cref{fig:fdof-metadata}). Regarding metadata and characterisation, an object becomes a \mt{FDO} when it is characterised by its informational value (using \mt{FDIO}), its materialisation format (using \mt{FDMO}) and is described by a \mt{FMR}.

\subsection{Ontological Considerations}
\label{sec:FDO-ontological-considerations}

The FDOF-CM has been designed with the requirement of making it as simple as possible, so that it can be understood in a wider context including all different types of web practitioners. However, it incorporates a number of ontological ideas and design patterns. 

The mechanism of type characterization in \cref{fig:fdof-attribution-and-identification,fig:fdof-characterisation} embeds the elements of the \textit{Four-Category Ontology} pattern (found in Foundational Ontologies such as UFO, DOLCE, GFO and BFO). Here we have: an \mt{Object} instantiates an \mt{Object Type} and is the \textit{bearer of a quality} that \textit{exemplifies} a given \mt{Quality Type}. Types here are defined as aggregations of \mt{Properties} (as in \textit{hyperintensional} type definitions). Actual quality individuals are represented by \mt{Attributions} that connect the bearer (the \mt{Object} at hand) with a projection of that quality as a \mt{Property Value} in a given \mt{Property Value Space}. This extends the classical Four-Category Ontology into the so-called \textit{Ontological Octagon} as proposed by \cite{guizzardi2012towards}. 

The mechanism in Fig, \ref{fig:fdof-metadata} that treats \mt{FAIR Metadata Records (FMRs)} as \mt{FDIOs} allows for the former to instantiate proper \mt{FAIR Digital Object Types} and hence be characterized by \mt{properties}, but also be described by other \mt{FMRs}, themselves \mt{FDIOs} that can instantiate \mt{FAIR Digital Object Types} characterized by \mt{Properties} and described by \mt{FMRs}, and so forth. This implements a minimum set of features available in more sophisticated multi-level theories such as \cite{fonseca2021multi}. 

The relation between \mt{Information Objects} and \mt{Media Objects} seem at first to reflect the relation of \textit{generic dependence} between content entities, e.g., \textit{The Aleph} by J.L.Borges (as a particular content with fictional individuals, locations and events) and a particular physical copy of that book standing on my table, i.e., the notions of \textit{Information Content Entity} and \textit{Information Bearing Entity} in the Information Artifact Ontology (IAO) \cite{smith2013iao}. However, notice that a \mt{Digital Media Object}, e.g., the JPEG bit sequence encoding of a photo taken at John's 30th birthday (an \mt{Digital Information Object}) can have multiple concretisations as \mt{Digital Copies}, i.e., actual patterns inhering in John's laptop, in Mary's mobile phone, in Paul's USB stick, etc, i.e., \textit{complex modes} \cite{guizzardi2005ontological}. Thus, FDOF-CM decouples encoding/formating entities from their possible multiple concretisations. In summary, FDIOs are generically dependent on FDMOs (the \textit{materialization} relation in \ref{fig:fdof-metadata}), which are themselves generically dependent on \mt{FAIR Digital Copies (FDCs)}. Our notion of \mt{Media Object} is akin to the notion of \textit{Information Structure Entity} in IAO; our notion of \mt{FDC} is akin to \textit{Information Bearing Entities }in IAO. \mt{FDCs} will be fully developed in future work.     

\section{FAIR Digital Object Framework Conceptual Model implementation in OWL}
\label{sec:fdof-evaluation}

% While FDOF aims at the systematisation of concrete implementations of the FAIR principles, and FDOF-CM provides semantic clarification about the concept of FAIR Digital Object, and its characterisation and identification methods, the application of both takes place in a variety of technology spaces. The semantic web space is a notable example of such, where OWL ontologies are extensively employed to provide semantic annotation of both data and metadata. In this section, we develop an OWL implementation of FDOF-CM considering the technical aspects of its context and making use of it in a selection of representative example use cases.

While FDOF can be applied to a variety of technology spaces, the Semantic Web is among the most significant ones, where OWL ontologies are extensively employed to provide semantic annotation of both data and metadata. The FAIR Digital Object Framework Ontology\footnote{The FDOF implementation in OWL is available at \url{https://w3id.org/fdof/ontology}.} (FDOF-OWL), is an OWL implementation of FDOF-CM. This implementation extends the \textit{gUFO} ontology \cite{almeida2019gufo}, a lightweight implementation of UFO for the semantic web, representing the \mt{Digital Object} and \mt{FAIR Digital Object} (\cref{fig:fdof-attribution-and-identification,fig:fdof-characterisation}) taxonomies as specialisations of gUFO's classes. In this section, we demonstrate the realisability of the FDOF-CM by implementing it in OWL and using one of the real use cases \footnote{The complete version of all use cases used in the validation of FDOF are  available at \url{https://w3id.org/fdof/fois23-paper}.}.

%\subsection{The FAIR Digital Object Framework in OWL}
%\label{sec:fdof-owl}

% The FAIR Digital Object Framework Ontology\footnote{The FDOF implementation in OWL is available at \url{https://w3id.org/fdof/ontology}.} (FDOF-OWL), is an OWL implementation of FDOF-CM. This implementation extends the \textit{gUFO} ontology \cite{almeida2019gufo}, a lightweight implementation of UFO for the semantic web, representing the \mt{Digital Object} and \mt{FAIR Digital Object} (\cref{fig:fdof-attribution-and-identification,fig:fdof-characterisation}) taxonomies as specialisations of gUFO's classes. 

% GUPRIs and IRIs

% While refining some aspects of FAIR's principles, FDOF opens the question of what makes a digital object an FDOF. While different FAIR applications require different levels of detail about the (meta)data they manipulate, a minimal property is involved in 

In FDOF-CM, one of the minimal requirements differentiating FDOs from regular digital objects is the identification by a GUPRI. However, the object identifiers in OWL are commonly used as IRIs when making statements about them. To differentiate objects from GUPRIs, allowing us to make statements about both independently, FDOF-OWL defines two distinct properties. The first, \mt{fdof:gupri}, is a datatype property which can be assigned any data value that represents the subject's GUPRI. The second, \mt{fdof:isIdentifiedBy}, is an object property whose range is an individualized \mt{fdof:Identifier} object. These two uses, presented in \cref{owl:fdof-gupri} through a dataset of Amazon's top-selling items, allow for independent statements about objects and GUPRIs, the first supporting coinciding GUPRIs and IRIs, and the second supporting statements about the GUPRIs themselves.

% @prefix : <https://w3id.org/fdof/fois23-paper/ex3/> .
\begin{lstlisting}[language=SPARQL,caption={Identification of FDOs in OWL.},label={owl:fdof-gupri}]
@prefix : <https://w3id.org/fdof/fois23-paper/ex1/> .
@prefix fdof: <https://w3id.org/fdof/ontology#> .
@prefix fdoft: <https://w3id.org/fdof/types#> .
@prefix dct: <http://purl.org/dc/terms/> .
@prefix dcat: <http://www.w3.org/ns/dcat#> .
@prefix schema: <https://schema.org/> .

:amazonTop50 fdof:gupri "https://w3id.org/fdof/fois23-paper/amazonTop50" ;
  fdof:isIdentifiedBy <https://w3id.org/fdof/fois23-paper/amazonTop50_identifier> .
\end{lstlisting}

% Metadata Records as named graphs

Next, FDOF-CM defines FAIR metadata records as datasets aggregating data items (\cref{fig:fdof-metadata,fig:fdof-attribution-and-identification}), information objects that record the attribution of values to digital objects. In the Semantic Web, data items coincide with the statements (triples) of an RDF/OWL file. To individualise and aggregate these data items, FDOF-OWL defines no data item concept but instead makes use of RDF's named graphs \cite{carroll2005named}. In \cref{owl:fdof-named-graph}, named graphs allow us to represent a set of statements about a \mt{dcat:Dataset} called \mt{:amazonTop50}, and refer to this set as \mt{:amazonTop50Metadata}. Other statements about \mt{:amazonTop50} could exist outside the named graph, but in this manner, we represent those aggregated in this metadata record, including the \mt{fdof:isMetadataOf} reference back to the metadata. Moreover, independent statements can be made about \mt{:amazonTop50Metadata}, e.g., a statement about its license as it could also differ from the license of \mt{:amazonTop50}.

% \noindent
% \begin{minipage}{\linewidth}
\begin{lstlisting}[language=SPARQL,caption={Named graphs representing FAIR metadata records.},label={owl:fdof-named-graph}]
{
 :amazonTop50Metadata rdf:type fdof:FAIRMetadataRecord ;
   fdof:gupri "https://w3id.org/fdof/fois23-paper/ex1/amazonTop50Metadata" ;
   fdof:hasInformationObjectType fdoft:Dataset ;
   dct:license <https://creativecommons.org/publicdomain/zero/1.0/> .
}

:amazonTop50Metadata {
  :amazonTop50Metadata fdof:isMetadataOf :amazonTop50 .
  :amazonTop50  rdf:type dcat:Dataset, fdof:FAIRDigitalInformationObject ;
    fdof:gupri "https://w3id.org/fdof/fois23-paper/ex1/amazonTop50" ;
    fdof:hasInformationObjectType fdoft:DatasetMetadaRecord ;
    dct:license <https://creativecommons.org/publicdomain/zero/1.0/> ;
    dct:issued "2020-10-01"^^xsd:date .
}
\end{lstlisting}
% \end{minipage}

% Defining FAIR Digital Object Types

As described in \cref{fig:fdof-characterisation}, FDOF includes a notion of FAIR information object types used to prescribe the characteristics required of a FAIR information object in a particular setting. In the context of OWL, these information object types are defined as subclasses of \mt{fdof:FAIRDigitalInformationObject}, and \mt{fdof:hasInformationObjectType}, a sub-property of \mt{rdf:type}, is used to connect the information object to the type characterising it as a FAIR (enough) FDO. In \cref{owl:fdof-named-graph}, \mt{:amazonTop50} and \mt{:amazonTop50Metadata} have relations to hypothetical FAIR information object types \mt{fdoft:Dataset} and \mt{fdoft:DatasetMetadata} which could set requirements beyond having a GUPRI, such as having explicit licensing information through \mt{dct:license} statements, and to include the publication date of a dataset through \mt{dct:issued}.

Finally, FDOF-OWL includes the object properties \mt{fdof:isMaterializedBy} and \mt{fdof:hasEncodingFormat}, both related to objects of type \mt{fdof:FAIRDigitalMediaObject}. The \mt{fdof:isMaterializedBy} property, connects an information object to the digital media object where it is materialized, and the \mt{fdof:hasEncodingFormat} property connects the digital media object to its characterising encoding format, much like the characterisation provided by \mt{fdof:hasInformationObjectType} to digital information objects as both instruct consumers on how to process subject digital object. In \cref{owl:fdof-media-object} we highlight statements about the media objects from our validation where \mt{:amazonTop50} is said to be materialized in a CSV file, while \mt{:amazonTop50Metadata} is materialized in the very Trig file where all these statements were declared.

% \noindent
% \begin{minipage}{\linewidth}
\begin{lstlisting}[language=SPARQL,caption={FAIR Media Objects in OWL.},label={owl:fdof-media-object}]
:amazonTop50  rdf:type dcat:Dataset, fdof:FAIRDigitalInformationObject .
  fdof:isMaterializedBy :amazonTop50Csv .

:amazonTop50Metadata rdf:type fdof:FAIRMetadataRecord ;
  fdof:isMaterializedBy :amazonTop50MetadataTrig .

:amazonTop50Csv rdf:type dcat:Distribution, fdof:FAIRDigitalMediaObject ;
  fdof:hasEncodingFormat <https://iana.org/assignments/media-types/text/csv> .

:amazonTop50MetadataTrig rdf:type dcat:Distribution, fdof:FAIRDigitalMediaObject ;
  fdof:hasEncodingFormat <https://iana.org/assignments/media-types/application/trig> .
\end{lstlisting}
% \end{minipage}

\begin{figure}[ht]
    \centering
    \includegraphics[width=\linewidth]{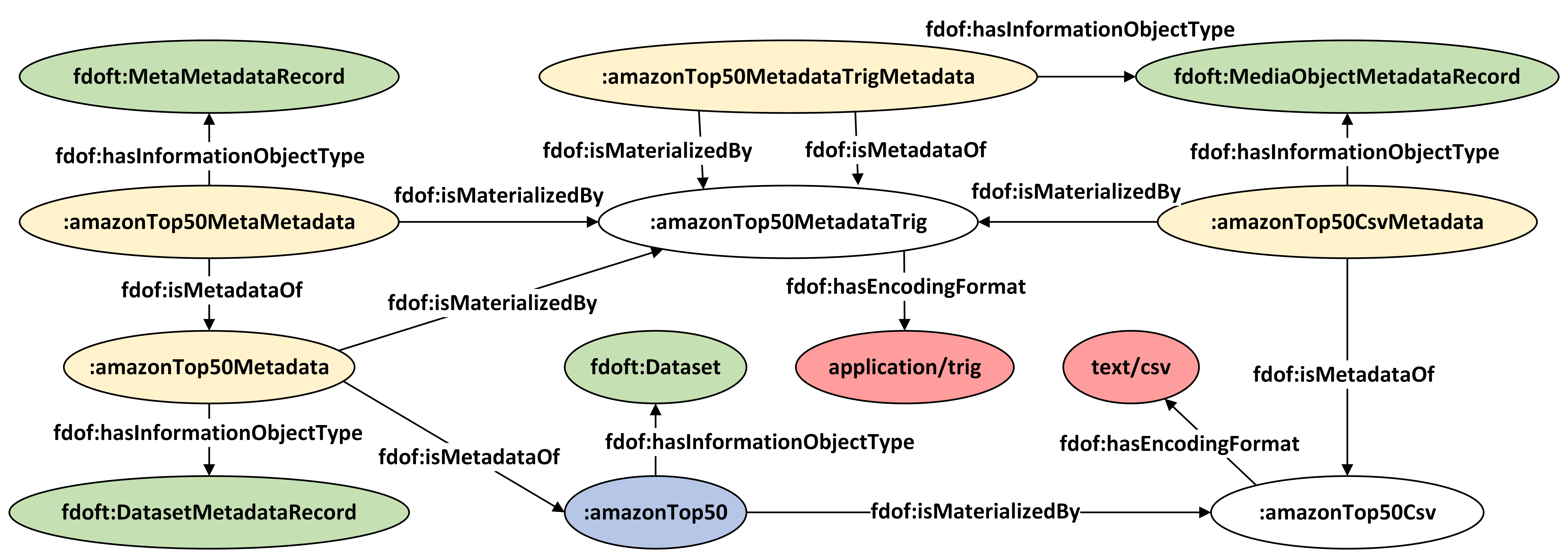}
    \caption{OWL dataset example in FDOF.}
    \label{fig:dataset-example}
\end{figure}

In \cref{fig:dataset-example} we summarise in a graph notation all digital objects of the dataset example used to validate the FDOF-CM. The complete case includes a real dataset, information objects for its metadata and meta-metadata, and information objects for the media objects materialising them and their own metadata. To improve the graph's readability, we also employ a colour coding of the FDOF types instantiated\footnote{Yellow represents FAIR metadata records; white represents FAIR media objects; blue represents FAIR information objects; red represents encoding formats (IANA media types in this case); and green represents domain information object types.}. FDOF-OWL has been validated through the representation of real use cases involving creative works (photos with multiple materialisations) and ontologies (including both gUFO and FDOF-OWL itself).

% \begin{lstlisting}[language=OWL,caption={Extending FDOF in OWL.},label={owl:fdof-extension}]
% :x123 rdf:type fdofo:DigitalObject ;
%   doi:doi "doi:123123123" 
%   fdof:isIdentifiedBy :doi_x123 ;

% doi:doi rdfs:subPropertyOf fdof:gupri ;
%     fdof:resolver <https://doi.org/> ;
%     fdof:persistancePolicy "P30Y"^^xsd:duration ;
%     fdof:managedBy <https://www.wikidata.org/wiki/Q11713409> .
    
% :doi_x123 rdf:type doi:Gupri . 
% \end{lstlisting}

% \subsection{Examples of FAIR Digital Objects}
% \label{sec:fdof-examples}

\section{Related Work}
\label{sec:related-work}

As aforementioned, the FDOF has been inspired by the notion of Digital Objects. In \cite{kahn_framework_2006} the authors introduce naming conventions to identify and located digital objects, a service to locate objects from their names and elements of a digital object access protocol. Two key entities are introduced: the obvious digital object and a repository. A digital object is defined as an instance of an abstract data type composed of two elements, data and key-metadata. The key-metadata includes the object's identifier and other information about the object. The paper also introduces the initial expected behaviours of the Repository Access Protocol (RAP), such as access and deposit digital objects.

The work on Digital Objects continued over the years, mainly under the scope of the DONA Foundation \cite{DONA}. A key result of this work is the Digital Object Architecture (DOA) \cite{DOA}. The DOA defines three core components, namely the identifier/resolution system, the repository system, and the registry system, as well as two protocols: the Digital Object Interface Protocol (DOIP)\cite{DOIP2} and the Digital Object Identifier Resolution Protocol (DO-IRP)\cite{DOIRP3}. The identifier/resolution system uses the  DO-IRP to create, update, delete, and resolve identifiers of digital objects. The repository system is responsible for supporting the management of digital objects, including the provision of access to the stored objects through their identifiers. The DOIP is used by the repository system and allows software applications to interact with digital objects. The registry system is a specialisation of repository system which stores the metadata of the digital objects rather than the digital objects themselves. In recent years, part of the work on DOs has been conducted in the context of the FAIR Digital Objects Forum \cite{FAIRDO}. The FDO Forum aims at providing a community structure to foster the advance in the specification and application of FAIR Digital Objects. A number of working groups have been set to address different aspects of FDOs. The work reported in the paper has been partially presented and tested in the FAIR Digital Object Semantics Group (FDO-SIG) of the forum.

The major differences between the work on the FDOF and the DO are that the latter has been designed from the ground up to follow the aspects of the FAIR principles that can be supported by technological infrastructures and applications, and the emphasis on semantic clarity by means of the FDOF-CM.

Regarding semantic models to describe digital objects, we relate our work with other initiatives such as Dublin Core Terms (DCT) \cite{DublinCoreTerms}, the Data Catalog Vocabulary (DCAT) \cite{DCAT2}, Schema.org, and the Metadata for Ontology Description and publication (MOD) \cite{MOD}. These models provide classes and relations that are used to represent entities in metadata records and to provide semantic annotations to data. The FDOF-CM does not aim at replacing them. Instead, as shown in the examples on \cref{sec:fdof-evaluation}, we reuse concepts from these models. However, in FDOF-OWL we provide greater expressivity for aggregating statements into well-defined metadata records, and we lay a foundation for the definition of more refined types of FAIR digital objects tailored for specific applications.
% In the implemented FDOF OWL ontology, whenever appropriate, we provide information that concepts in the FDOF are related to concepts in these models.

\section{Final Remarks and Future Work}
\label{sec:conclusions}

In this paper, we reported the current state of the work on the conceptual model of the FAIR Digital Object Framework. In particular, we describe how FAIR Digital Objects are identified, how they can be classified through the typing system, and how the model follows the FAIR principles also in regard to metadata.

We demonstrated the utility of the conceptual model in the context of the Semantic Web by generating an OWL ontology and using this ontology to describe the objects in representative real use cases. We argue that characterising objects by their informational value, encoding format, and related entities, provides valuable information contributing to semantic clarity. We acknowledge that this approach not only brings these benefits but also requires the existence of more instances with their consequent properties and relations, which may increase complexity. However, it is not expected that these instances, their identifiers and some of their properties would be manually created. Supporting systems can be implemented to automate this process, generating automatically these instances, identifying and setting some of their properties and, thus, making the adoption of the FDOF easier.

In terms of future work, the FDOF-CM and FDOF-OWL, presented in \cref{sec:fdof-ontology,sec:fdof-evaluation}, lay the foundation for the systematic characterisation and representation of FAIR Digital Objects according to FDOF. First of all, the mechanisms of representation of information objects, their metadata, and their materialisations involve a lot of nuances that have been left out of our scope, and enhancing this coverage also includes support to non-digital objects, a subject of great relevance. Second, we focused this paper on the characterisation of FDOs based on the attributions aggregated into FAIR metadata records, leaving out the characterisation of operation types present in FDOF. FDOF's operation types are fundamental to explaining to both human and software agents how and when to process information objects, in a finer level of detail than what is possible through their materialisations and encoding types. For example, in the case of the Digital Object Architecture and DOIP support the existence of basic and extended operations. While basic operations define a set of operations that every DOIP service must support, extended operations allow different services to implement extra operations that may be offered. Third, we foresee the possibility of employing automated verification techniques, such as SHACL, for the definition of \mt{FAIR Information Object Types} and the verification of their instances. This should guarantee that these instances sufficiently adhere to the FAIR principles according to the requirements defined by the application extending FDOF-OWL.

% by Luiz
% Although the example presented in \cref{sec:fdof-evaluation} shows the current expression capabilities of the conceptual model, we have already identified a number of directions to evolve the work. The description of the conceptual model and the use case examples were based on objects in the digital realm as it was the scope for this paper. However, the work on evolving the model to accommodate the possibility of FDOs referring to physical objects or FAIR Metadata Records describing physical objects is part of our next steps.

% by Luiz
% Another area of interest to extend the current FDOF-CM relates to operations. The Digital Object Architecture and, in particular, the DOIP support the existence of basic and extended operations. While basic operations define a set of operations that every DOIP service must support, the extended operations allow different services to implement extra operations that may be offered. In the FDOF, we intend extend the concept of \mt{Operation Type} depicted in \cref{fig:fdof-characterisation} to define operations related to objects types. The goal is to allow the definition of which operations a given object type entails, which operations are supported by services managing FDOs and which operations are allowed considering who wants to interact with the object. 

Finally, the work on the FDOF conceptual model occurs in parallel with the work on the rest of the framework. Reports on the whole framework, on the definitions of the protocols and expected behaviours of the participant applications and services, as well as prototypes and demonstrators, are currently underway.

% TODO (Someone) Talk about variants and versions.

\bibliographystyle{vancouver}
\bibliography{references}
\end{document}